\documentclass[epj]{svjour}
\usepackage{graphicx}

\begin{document}

\title{ Evidence for creation of strong electromagnetic fields
 in relativistic heavy-ion collisions}

\author{V. Toneev\inst{1} \and O. Rogachevsky\inst{1} \and V. Voronyuk \inst{1,2}}%
\institute{1. Joint Institute for Nuclear Research, Dubna, Russia\\
2. Bogolyubov Institute for Theoretical Physics, Kiev, Ukraine}

\abstract{It is proposed to identify a strong electric field
created during relativistic collisions of asymmetric nuclei via
 observation of pseudorapidity and transverse momentum distributions
 of hadrons  with the same mass but opposite charges. The detailed
 calculation results for the directed flow within the Parton-Hadron String
 Dynamics model are given for Cu-Au interactions at the NICA collision energies of
 $\sqrt{s_{NN}}=9$ and $5$ GeV. The separation effect is observable at 9 GeV as
 clearly as at 200 GeV.}

 \maketitle

It was demonstrated in Ref.~\cite{SIT9} that charged particles
created in relativistic heavy-ion collisions may produce extremely
strong electromagnetic field strength. In particular, in
peripheral Au-Au collisions at the energy $\sqrt{s_{NN}}=$200 GeV
in the very initial interaction state this strength reaches
$|eB_y| \sim 5 m_\pi^2 = 5\cdot 10^{18}$ Gauss,  which exceeds every
value reachable in the earth conditions and even may be higher
than the fields created in magnetars. However, the subsequent
analysis of Au+Au collisions in the range up to the top RHIC
energy has observed no visible effect of the strong electromagnetic 
interactions on global characteristics,
in particular, for such a sensitive quantity as the elliptic flow.
The
\begin{figure}[th]
\centering
\includegraphics[width=0.5\textwidth]{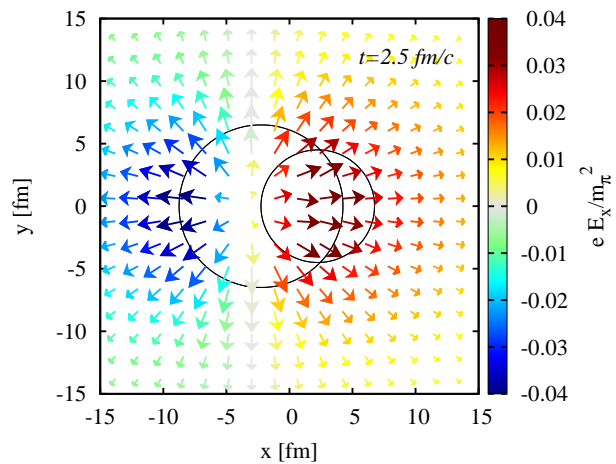}
\caption{Electric field generated in the transverse plane of Cu+Au
collisions at $\sqrt{s_{NN}}=$9 GeV, $b=$4.5 fm and $t=$2.5 fm/c.
The direction of arrows indicates the field direction projected
onto the reaction plane and the length is proportional to the
electromagnetic strength shown in color. }
 \label{diagCuAu9-E}
\end{figure}
reason of that is not a very short interaction time of the system,
as could be naively expected, but rather a compensation effect
between electric and magnetic components of acting electromagnetic
forces, as found in Ref.~\cite{VTC11}. Thus, the question of
wether so strong electromagnetic fields are really created in
high-energy nuclear collisions remains open.

\begin{figure}[th]
\centering
\includegraphics[width=0.45\textwidth]{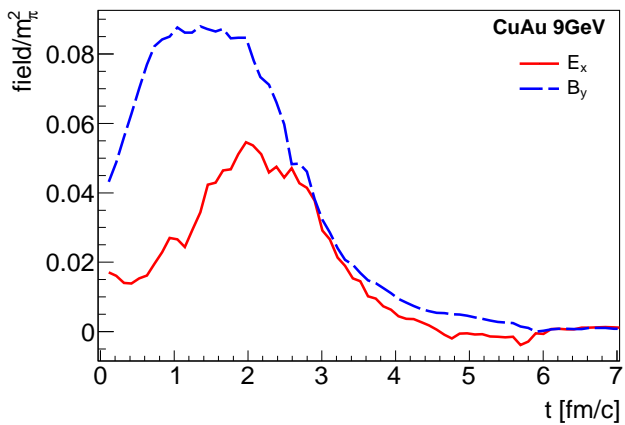}
\caption{The time dependent components of the average electric 
$E_x$ and magnetic $B_y$ field generated in the center of Cu+Au
collision at $b=$4.5 fm and $\sqrt{s_{NN}}=$9 GeV. }
 \label{EBt-9}
\end{figure}

Recently, it has been found that such compensation is absent in
asymmetric  relativistic collisions due to the difference in the number of
protons in colliding nuclei~\cite{VTV14,HHH14}. In particular, in
Cu+Au(200 GeV) collisions the directed flow (the first flow
harmonic $v_1 (\eta,p_t)$) exhibits the electric charged particle
dependence. As follows from the results presented in
Fig.~\ref{diagCuAu9-E},  the electric field in the central region
of the overlapping area has a specific tendency to go from
Au to Cu and this effect can be
observed at the nominal NICA energy $\sqrt{{s_{NN}}}=$9 GeV, too.
Indeed, the strength of the induced electric field is
strongly asymmetric within the region of overlapping nuclei, which
may result in observable asymmetry of hadrons with opposite
charges.
\begin{figure}[th]
\centering
\includegraphics[width=0.4\textwidth]{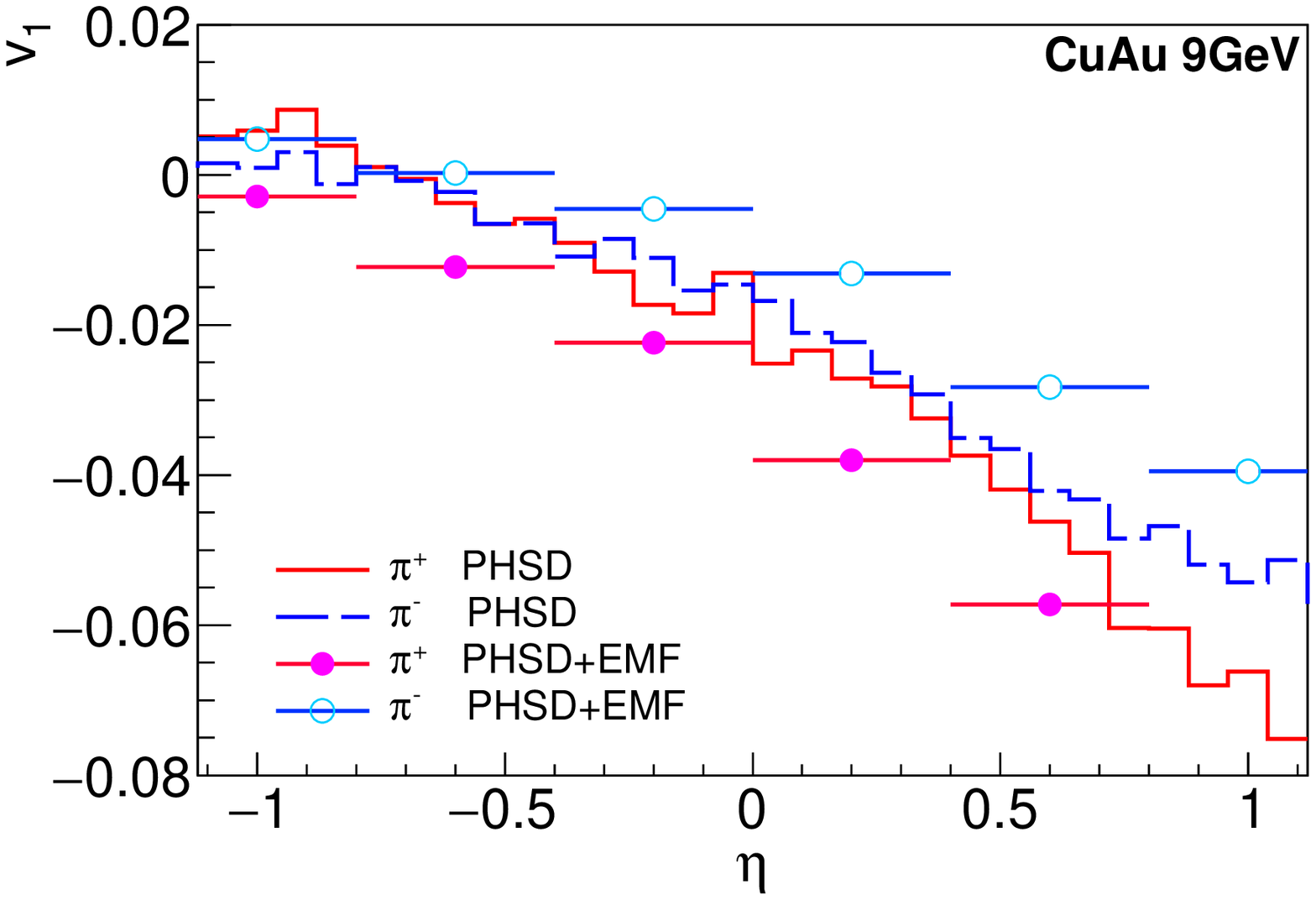}\\
\vspace{5mm}
\includegraphics[width=0.4\textwidth]{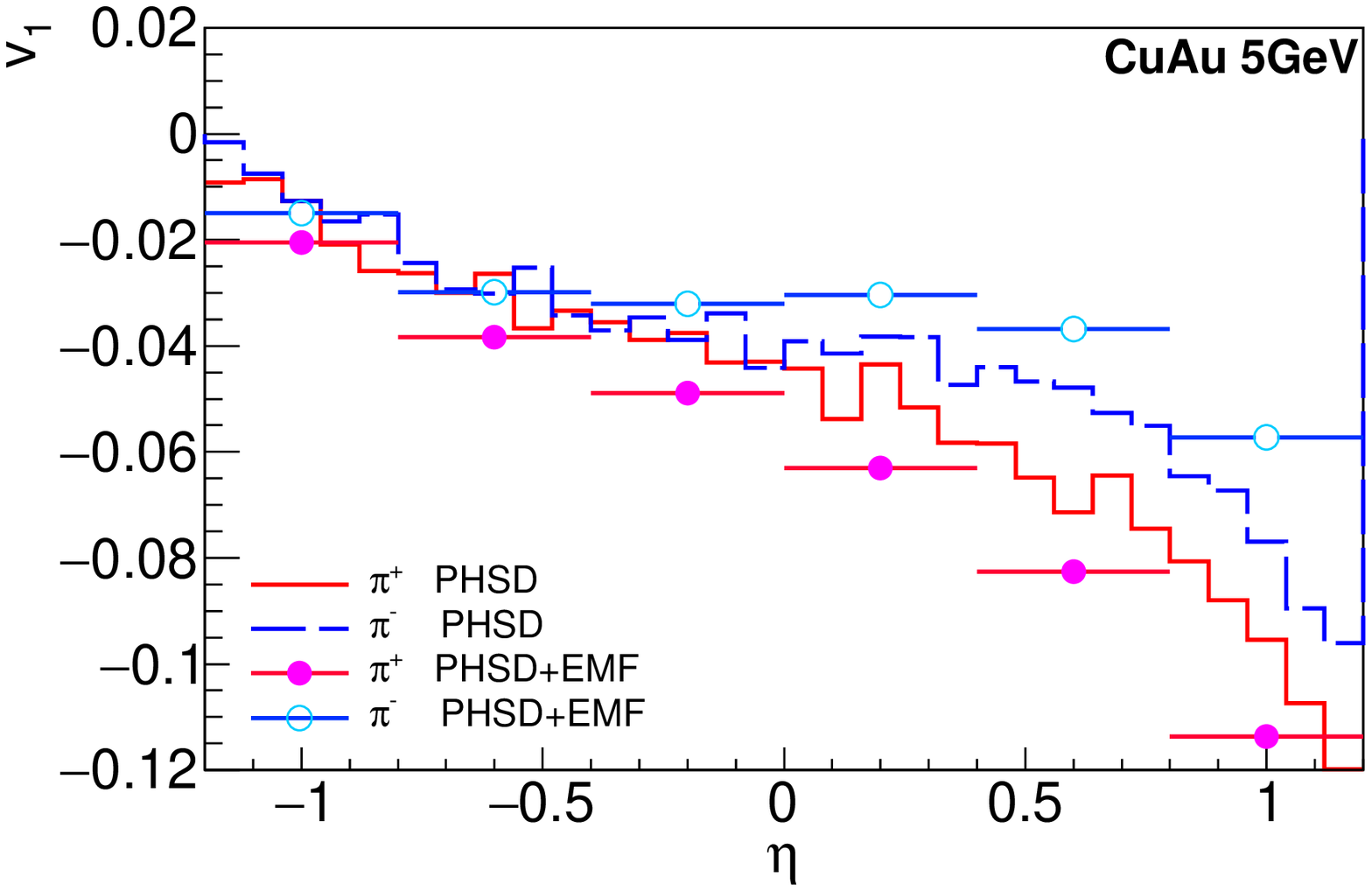}
\caption{Pseudorapidity distribution of positive and negative
pions created in Cu+Au collisions at $\sqrt{s_{NN}}=$9 (top figure) and 5 (bottom)
GeV and the impact parameter interval 4.4-9.5 fm.
Histograms are the result of the standard PHSD transport approach,
points are the results where electromagnetic force is additionally included. }
 \label{v1eta}
\end{figure}

As noted in \cite{VTC11}, the electromagnetic field (EMF) is formed
predominantly by spectators in the early time during the mutual passage
of the two colliding nuclei. Since the number of spectators increases
with the impact parameter $b$, the magnetic field should also
increase in more peripheral collisions and decrease gradually
with $b$. In addition, the directed flow $v_1$ is
getting larger when one proceeds to more peripheral collisions.
Thus, this experiment is more promising at large impact parameters.

\begin{figure}
\includegraphics[width=0.45\textwidth]{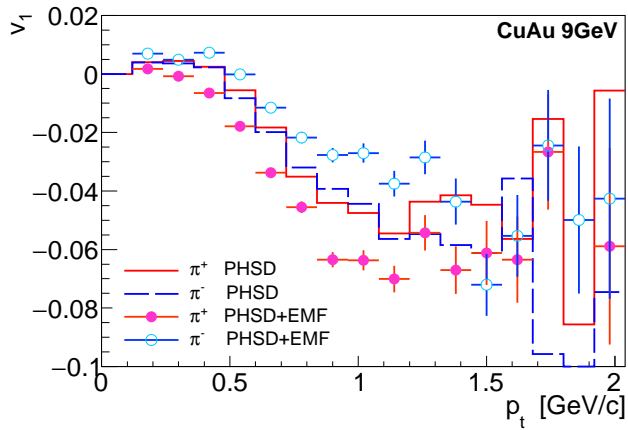}\vspace{5mm}
\includegraphics[width=0.45\textwidth]{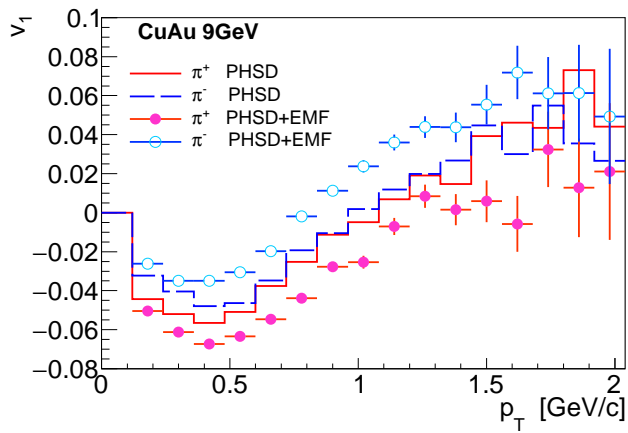}
\caption{Charge-dependent  $p_t$ distributions of  pions from
asymmetric Cu+Au collisions at $\sqrt{s_{NN}}=$9 GeV for backward
(upper panel) and forward (bottom panel) emission.  The parameters used
and notation - as in Fig.~\ref{v1eta}. }
 \label{v1pt}
\end{figure}

As is seen from Fig.~\ref{EBt-9}, the average strength of the dominant
components of electric $\langle E_x\rangle$ and magnetic  $\langle B_y\rangle$ 
fields reaches in the time interval 1-2 $fm/c$ maximal values  
$\langle eE_x\rangle\approx 0.050 m^2_{\pi}$ and magnetic
$\langle eB_y\rangle\approx 0.085 m^2_{\pi}$. Other components are
practically negligible.  These values are obtained in the spatial cylinder 
with the radius $R=2$ fm and the length $|z|<2.5/\gamma$ fm passing through 
the center of colliding nuclei. The maximal average energy density reached in
the same cylinder is about 1.6 and 0.9 $GeV/fm^3$ for the collision energy 9 
and 5 GeV, respectively, that justifies the use of the
partonic version of the string dynamic model.

In this respect, on the basis of the Parton Hadron String Dynamics
model~\cite{CB-PHSD} we calculated various characteristics of
asymmetric Cu+Au collisions at the NICA energy $\sqrt{s_{NN}}$ =9
and 5 GeV. As is seen from Fig.~\ref{v1eta}, without accounting for the
created EMF  the $\eta$ distributions for
$\pi^+$ and $\pi^-$ are very close to each other (difference is
coming only from different mean multiplicity of these pions), but
the inclusion of the electromagnetic field results in a sizable
separation of these distributions. Note that the detector
acceptance is taken into account here. At lower energy 5 GeV the
separation effect is  weaker: the difference between $v_1(\eta)$
distributions calculated with and without the EMF is within statistical
error bars.

The transverse momentum distributions of the directed flow $v_1$ for
pions created at $\sqrt{s_{NN}} =$9 GeV are presented in Fig.~\ref{v1pt}.
It is remarkable that
$v_1(p_t)$  differs in the backward ($\eta<0$) and forward
($\eta>0$) directions. For $\eta<0$ (the Au side) the low momentum
($p_t<$1 GeV/c) pions are dominated while in the opposite
direction the $p_t$ distribution is an increasing function with
some minima at $p_t\sim 0.5$ GeV/c. In both cases the charge
separation is well visible in Fig.~\ref{v1pt} and it is getting
larger for higher values of $p_t$.
\begin{figure}
\includegraphics[width=0.47\textwidth]{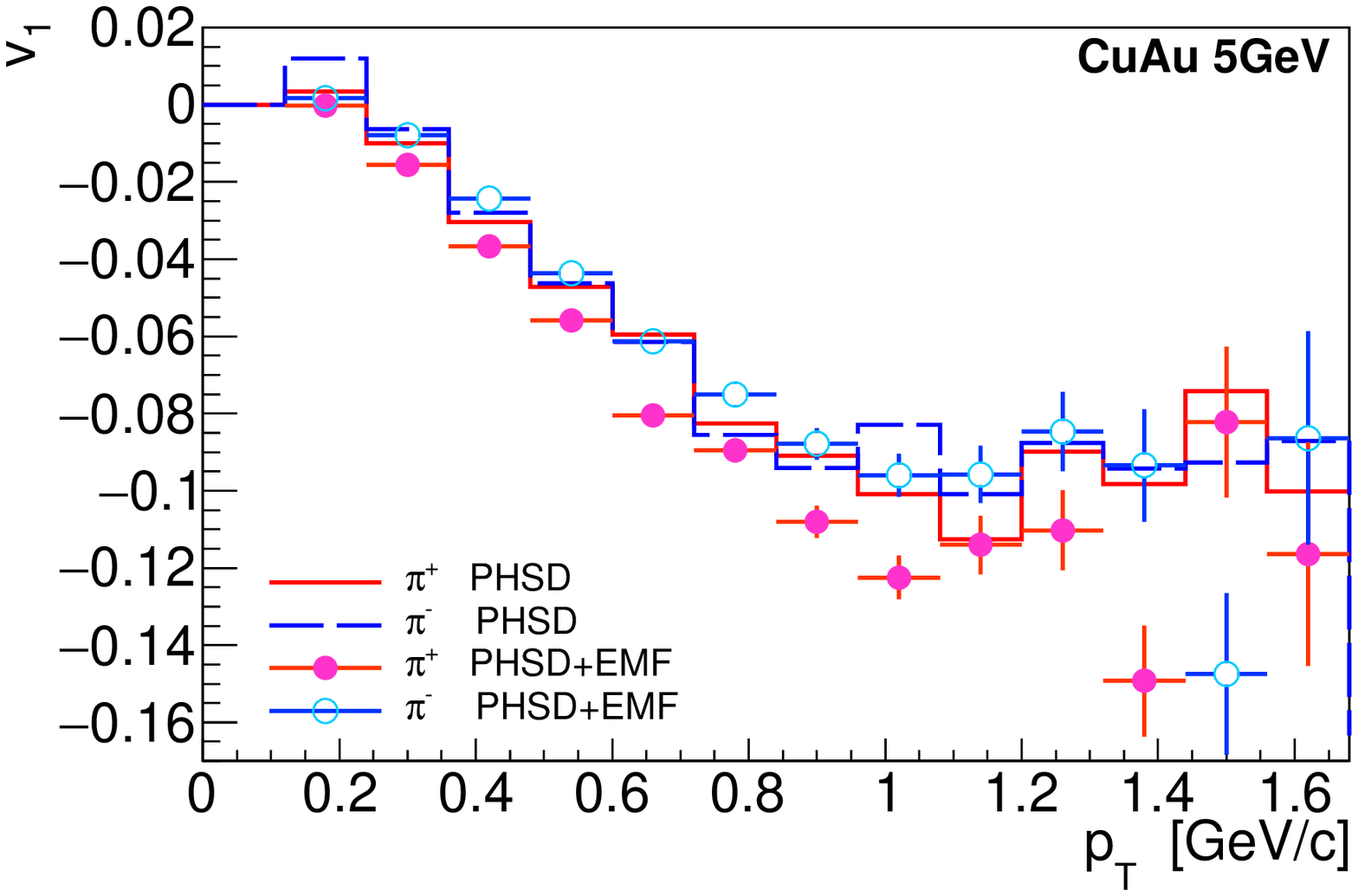}\vspace{5mm}
\includegraphics[width=0.47\textwidth]{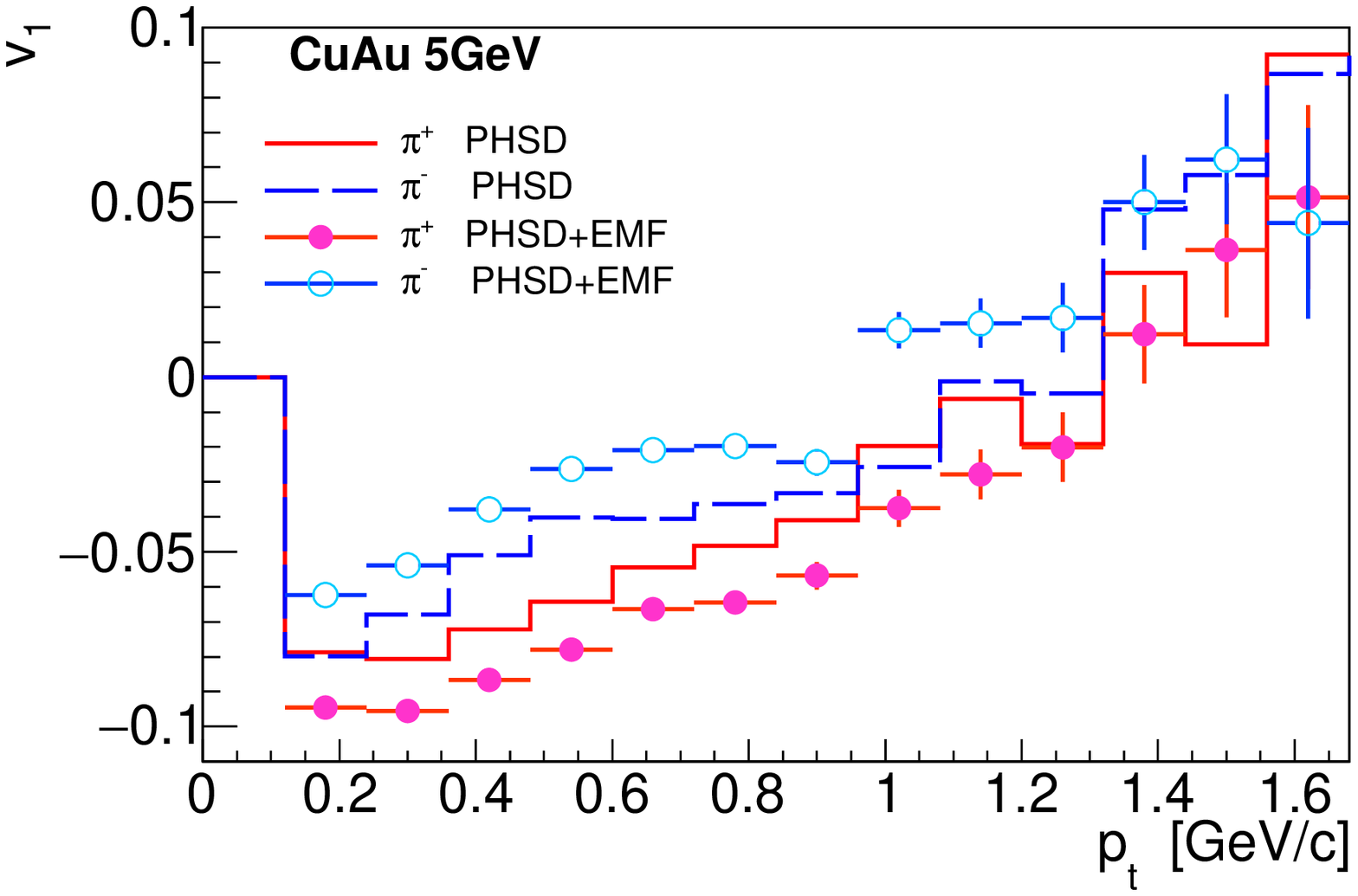}
\caption{Charge-dependent  $p_t$ distributions of  pions from
asymmetric Cu+Au collisions at $\sqrt{s_{NN}}=$5 GeV for backward
(upper panel) and forward (bottom panel) emission. Notation - as in Fig.~\ref{v1eta}. }
 \label{v1pt5}
\end{figure}

The pion transverse momentum distributions  for the collision energy 5 $GeV$ are
shown in Fig.~\ref{v1pt5}. It is clearly seen that though the functional behavior of $v_1(p_t)$
is quite similar to that in Fig.~\ref{v1pt}, the magnitude of the pion directed flow $v(p_t)$ is
 suppressed  at lower energy resulting in smaller separation effect. It is noteworthy
that the magnitude of $v_1$ is a weakly increasing function above $\sqrt{s_{NN}}=$9 GeV and
practically coincides with appropriate values at the collision energy 200 GeV \cite{VTV14}.

Similar effects are observed for $K^+$ and $K^-$ mesons. However, for a
proton-antiproton pair this is not the case since the strong
interaction of $p$ and $\bar{p}$ is quite different.

The particularity of  the NICA energy range  is that the
particle creation  occurs at a high baryon density or a large
baryonic chemical potential. In addition, the electric charge chemical potential
is also quite important since we are interested in hadrons with
opposite electric charges. Future study of these effects,
especially in the partonic phase, is of great interest.

 Thus, we propose to measure the pseudorapdity $v_1(\eta)$ and
transverse momentum $v_1(p_T)$ distributions  of  the directed
flow for identified pairs of hadrons (at least for $\pi^+, \pi^-$
and $K^+, K^-$ mesons) in the NICA energy range, which could first
evidence a new physical effect -- the formation of extremely
strong electromagnetic fields in relativistic heavy-ion
collisions. The top NICA energy is favorable for such measurements.
The realization of this experiment implies that two  ions
of different kinds may interact in the collider creating a
very high electromagnetic field. As an estimate
shows, the planned luminosity $L=10^{27}cm^{-2}s^{-1}$ resulting
in the collision rate $(\sigma L \epsilon) =$500 event/s makes
feasible this experiment at the NICA collider.

\end{document}